\documentclass[12pt]{article} \topmargin= 0.3cm \textwidth 16.5cm
=0cm \headsep =0cm
\usepackage {latexsym}
\usepackage {graphicx}

\begin{document}

\title{Gauss-Bonnet gravity, brane world models, and non-minimal coupling}
\author{K. Farakos \footnote{kfarakos@central.ntua.gr} and P. Pasipoularides  \footnote{paul@central.ntua.gr} \\
       Department of Physics, National Technical University of
       Athens \\ Zografou Campus, 157 80 Athens, Greece}
\date{ }
       \maketitle

\begin{abstract}
We study the case of brane world models with an
additional Gauss-Bonnet term in the presence of a bulk scalar field
which interacts non-minimally with gravity, via a possible
interaction term of the form $-\frac{1}{2} \xi R\phi^2$. The
Einstein equations and the junction conditions on the brane are
formulated, in the case of the bulk scalar field. Static solutions
of this model are obtained by solving numerically the Einstein
equations with the appropriate boundary conditions on the brane.
Finally, we present graphically and comment these solutions for
several values of the free parameters of the model.
\end{abstract}

\section{Introduction}

Hidden extra dimensions is a way to extend four dimensional
conventional physics. According to the Kaluza Klein picture, extra
dimensions are compactified to a very small length scale (naturally
the Planck scale), and as a result space time appears to be
effectively four dimensional, insofar as low energies are concerned.
An alternative picture, which explains why extra dimensions are
hidden, is the brane world scenario. In this case all ordinary
particles are assumed to be localized on a three-dimensional brane
(our world) which is embedded in a multidimensional manifold (bulk),
while gravitons are free to propagate in the bulk \cite{Akama,Shap}.
The brane world scenario has attracted great attention, as it
predicts interesting phenomenology even at the TeV scale \cite{ADD},
and puts on a new basis fundamental problems such as the hierarchy
and the cosmological constant problem. According to the nature of
the extra dimensions, there are two standard brane world models with
gravity: a) in the case of flat extra dimensions we have the
ADD-model \cite{ADD}, and b) in the case of warped extra dimensions
we have the first and second Randall-Shundrum models \cite{Ran}.

Brane Models with bulk scalar fields have also been considered. In
Ref. \cite{AH} the existence of a bulk scalar field serves as a way
to resolve the cosmological constant problem (see also \cite{Rub}
and references therein). However, the solutions of the model suffer
from naked singularities, and this implies a fine-tunning for the
five dimensional cosmological constant. An alternative role of the
scalar field is presented in Ref. \cite{Keh}. In this case the
scalar field creates the brane dynamically, by forming a kink
topological defect toward the extra dimension. Also, nontopological
defects appear for a special periodic form of the potential
\cite{nontop}.

In Refs. \cite{KF,FP} we have studied brane world models with a bulk
scalar field non-minimally coupled with gravity, via a possible
interaction term of the form $-1/2 \xi R\phi^2$. Our motivation was
to construct brane models which incorporate a layer-phase mechanism
\cite{Shif,KA,lat,Rumm} for the localization of the ordinary
particles on the brane. Similar models for phenomenological reasons
have also been consider in \cite{Toms,Dav}.

Particularly, in Ref. \cite{KF}, by solving numerically the Einstein
equations in the case of a non-minimally coupled scalar field, we
have obtained three classes of static solutions with different
features, in appropriate regions of the free parameters of the
model. The numerical approach we apply in \cite{KF} is suitable for
the general case of the potential. However, we have focused to the
case of a massless scalar field assuming the simplest generally
accepted form of the potential: $V(\phi)=\lambda \phi^4$.

An analytical study of the same brane model, with a nonminimally
coupled scalar, can be found in Ref. \cite{Tam}. In this paper
analytical solutions have been found by choosing suitably the
potential for the scalar field. For cosmological implications of
these models with the nonminimal coupling you can see \cite{Tamc}.

Furthermore, 5D models in the framework of Brans-Dicke theory have
been also studied. Particularly, in Ref. \cite{bd}, static
analytical solutions are constructed for a special class of
potentials, and the stabilization of the extra dimension is
discussed.

An alternative way to go beyond the  Randall-Sundrum model, is to
add a Gauss-Bonnet term to the gravity action, which, in the case of
$D\geq 5$, has the property to keep equations of motion to second
order. In the case of $D=4$ the Gauss-Bonnet term is topological and
does not contribute to the classical equations of motion, however it
has nontrivial consequences at the level of the conserved currents
of the theory \cite{Cur}. It is worth noting that, a Gauss-Bonnet
term arises as the leading order quantum correction in the case of
the heterotic string theory. The Randall-Sundrum model with an
additional Gauss-Bonnet term has been considered in
\cite{GBRS,New0,spec,Odi}. Brane models with scalar fields and
Gauss-Bonnet gravity has been also examined in
\cite{NM,New,Charm,MG}, while for cosmological implications see for
example \cite{CH,EP}.

In this paper we will study brane world models with Gauss-Bonnet
gravity in the presence of a non minimally coupled scalar field. The
motivation is to examine how Gauss-Bonnet gravity affects the three
classes of solutions we obtained in our previous work \cite{KF}.

\section{Gauss-Bonnet gravity, brane world models, and non-minimal coupling}

In this section we will study brane world models with an action of
the form:
\begin{equation}
S=\int d^5 x\left({\cal L}_{RS}+{\cal L}_{GB}+{\cal L}_{\phi}\right)
\end{equation}
where $d^{5}x=d^{4}x dz$, and $z$ parameterizes the extra dimension.

As we see the above mentioned lagrangian consists of three terms.
The first term corresponds to the lagrangian of the RS2-model:
\begin{equation}
{\cal L}_{RS}=\sqrt{|g|}\left (\frac{1}{2}
R-\Lambda-\sigma\delta(z)\frac{\sqrt{|g^{(brane)}|}}{\sqrt{|g|}}\right)
\end{equation}
where $\Lambda$ is the five-dimensional cosmological constant and
$\sigma$ is the brane tension. In addition $R$ is the
five-dimensional Ricci scalar, $g$ is the determinant of the
five-dimensional metric tensor $g_{MN}$ ($M, N=0,1,...,4$), and
$g^{brane}$ is the determinant of the induced metric on the brane.
We adopt the mostly plus sign convention \cite{R2}.

The second term is the well-known lagrangian for Gauss-Bonnet
gravity:
\begin{equation}
{\cal L}_{GB}=\frac{b}{2}
\left(R^2-4R_{MN}R^{MN}+R_{MNKL}R^{MNKL}\right)
\end{equation}
where we have introduced a new coupling b in conection with the
Gauss-Bonnet term. Assuming that this term arises from string theory
(tree level) the coupling b takes only positive values. However, in
this parer we will consider that the range of b covers all the real
axis.

The scalar field part of the lagrangian is
\begin{eqnarray}
{\cal L}_{\phi}&=&\sqrt{|g|}\left(-\frac{1}{2}g^{M N} \nabla_{M}\phi
\nabla_{N}\phi-\frac{1}{2}\xi \phi^2 R -V(\phi)\right)
\end{eqnarray}
where the potential is assumed to be of the standard form
$V(\phi)=\lambda \phi^{4}$.

The Einstein equations, which correspond to the action of Eq. (1)
are
\begin{eqnarray}
G_{MN}-\frac{b}{2}H_{MN}+\Lambda\; g_{MN}+\sigma
\delta(z)\frac{\sqrt{|g^{(brane)}|}}{\sqrt{|g|}}g_{\mu\nu}\delta_{M}^{\mu}\delta_{N}^{\nu}=T^{(\phi)}_{MN}
\end{eqnarray}
where the energy momentum tensor for the scalar field is
\begin{equation}
T^{(\phi)}_{MN}=\nabla_{M}\phi\nabla_{N}\phi-g_{MN}[\frac{1}{2}g^{P\Sigma}\nabla_{P}\phi\nabla_{\Sigma}\phi+V(\phi)]-2
\xi \nabla_{M}\nabla_{N}\phi+2 \xi g_{MN}\Box \phi+ \xi \phi^2
G_{MN}
\end{equation}
and $H_{MN}$ is the Lanczos tensor
\begin{equation}
H_{MN}={\cal L}_{GB}\; g_{MN}-4 R R_{MN}+8
R_{ML}R^{L}_{\:\:N}+8R_{MKNL}R^{KL}-4R_{MKLP}R_{N}^{\;\:\:KLP}
\end{equation}

The equation of motion for the scalar field is
\begin{equation}
\Box \phi-\xi \phi R-V'(\phi)=0
\end{equation}
The above equation is not independent from the Einstein equations
(5), as it is equivalent to the conservation equation $\nabla^M
T^{(\phi)}_{MN}=0$, where $T^{(\phi)}_{MN}$ is given by Eq. (6).

We are looking for static solutions of the form
\begin{equation}
ds^{2}=a^{2}(z)(-dx_{0}^{2}+dx_{1}^{2}+dx_{2}^{2}+dx_{3}^{2})+dz^{2},
\quad \phi=\phi(z)
\end{equation}
From the Einstein Equations (Eq. (5)) we obtain two independent
equations:
\begin{eqnarray}
&&G_{ii}-\frac{b}{2}H_{ii}+\Lambda\; g_{ii}+\sigma \delta(z)g_{ii}=T^{(\phi)}_{ii}\\
&&G_{zz}-\frac{b}{2}H_{zz}+\Lambda\; g_{zz}=T^{(\phi)}_{zz}
\end{eqnarray}
where $i=0,1,2,3$.

If we set $a(z)=e^{A(z)}$ and use Eqs. (9),(10) and (11) we get
\begin{eqnarray}
F_1=0:&&3(1-\xi\phi^{2}(z))\left(A''(z)+2 A'(z)^2\right)-12 b
A'(z)^4-12 b A'(z)^2 A''(z)+\Lambda\\&&+(\frac{1}{2}-2
\xi)\phi'(z)^2+V(\phi(z))-2 \xi \phi(z) \phi''(z)-6 \xi A'(z)\phi(z)
\phi'(z)+\sigma \delta(z)=0\nonumber
\end{eqnarray}
\begin{eqnarray}
F_2=0:6(1-\xi\phi^{2}(z)) A'(z)^2-12 b
A'(z)^4+\Lambda-\frac{1}{2}\phi'(z)^2+V(\phi)-8 \xi A'(z)\phi(z)
\phi'(z)=0
\end{eqnarray}
From Eq. (9) for the scalar field we obtain
\begin{eqnarray}
F_3=0:\quad -\phi''(z)-4 A'(z) \phi'(z)-\xi\left(8
A''(z)+20A'(z)^2\right)\phi(z)+V'(\phi)=0
\end{eqnarray}

Note that Eq. (14) can be found from the combination
$-4A'(z)(F_1-F_2)+F_2'=0$ of  Eqs. (12) ($F_1=0$) and (13)
($F_2=0$). As Eqs. (12),(13) and (14) are not independent, the
static solutions of the model can be obtained by integrating
numerically the second order differential equations (12) and (14).
The first order differential equation (13) is an integral of motion
of Eq. (12) and (14), and acts as a constraint between $A(z)$,
$\phi(z)$ and their first derivatives.

In particular we choose to solve numerically the following second
order differential equations:
\begin{eqnarray} &&3(1-\xi\phi^{2}(z))
A''(z)-12 b A'(z)^2 A''(z)+(1-2 \xi)\phi'(z)^2-2 \xi \phi(z)
\phi''(z)\nonumber \\&+&2 \xi A'(z)\phi(z) \phi'(z)+\sigma
\delta(z)=0
\end{eqnarray}
\begin{equation}
-\phi''(z)-4 A'(z) \phi'(z)-\xi\left(8
A''(z)+20A'(z)^2\right)\phi(z)+V'(\phi)=0
\end{equation}
Eq. (15) is obtained as the difference between Eqs. (12) and (13).

This system of differential equations is quite complicated, so we
will not look for analytical solutions. On the other hand we will
try to solve it numerically. For the numerical integration ($z\geq
0$) of Eqs. (15) and (16) it is necessary to know the values of
$A(0)$, $\phi(0)$, $A'(0^+)$ and $\phi'(0^+)$. These values are
determined by the junction conditions (see Eqs. (17) and (18) below)
and the constraint Eq. (13).

The junction conditions on the brane are obtained from Eq. (15) and
(16)
\begin{eqnarray}
&&6(1-\xi\phi^{2}(0)) A'(0^+)-8 b A'(0^+)^3-4\xi
\phi(0)\phi'(0^+)+\sigma=0\\
&&\phi'(0^+)+8\xi A'(0^+) \phi(0)=0
\end{eqnarray}
if we take into account that $A'(z)^2 A''(z)=1/3(A'(z)^3)'$. For
details see \cite{FP,NM,Charm,New0}.

The values of $\phi(0)$, $A'(0^+)$ and $\phi'(0^+)$ can be found by
solving numerically the system of algebraic equations (17), (18) and
equation (13) (if we perform the replacement $z=0$ in equation
(13)).

\section{Numerical Results}

The solutions (or the functions $A(z)$ and $\phi(z)$)) are obtained
by solving numerically the system of second order differential
equations (15) and (16). The boundary conditions on the brane
$\phi(0)$, $A'(0^+)$ and $\phi'(0^+)$ are obtained by solving
numerically the algebraic system of Eqs. (17), (18) and (13). Note
that $A(0)=0$, as $a(0)$ is normalized to unity, and
$a(z)=e^{A(z)}$.

We would like to emphasize that the parameter space of the model
includes five parameters $\xi,\lambda,\Lambda,\sigma,b$. Before we
proceed with the numerical analysis, it is convenient to make the
parameters dimensionless by performing the transformation
$z\rightarrow kz$. In order to use the notation of the RS-model,
without the scalar field, we choose $k=\sqrt{-\Lambda/6}$ (note that
in this paper we have examined mainly the case of a negative five
dimensional cosmological constant, or $\Lambda<0$). Now the number
of independent parameters is reduced to four, namely: $\;
\xi,\;\hat{\lambda}=\lambda/k^2,\;\hat{\sigma}=\sigma/k,\;\hat{b}=bk^2$.
After this, it is obvious that the relevant scale is the bulk
cosmological constant.

Of course, a complete investigation of a so extensive parameter
space, with four independent parameters, is beyond the scope of this
paper. Thus, we will restrict our study to the classes of numerical
solutions we obtained in \cite{KF}. We aim to investigate how
Gauss-Bonnet gravity affects these solutions.

\subsection{Conformal Coupling}

\begin{figure}[h]
\begin{center}
\includegraphics[scale=1,angle=0]{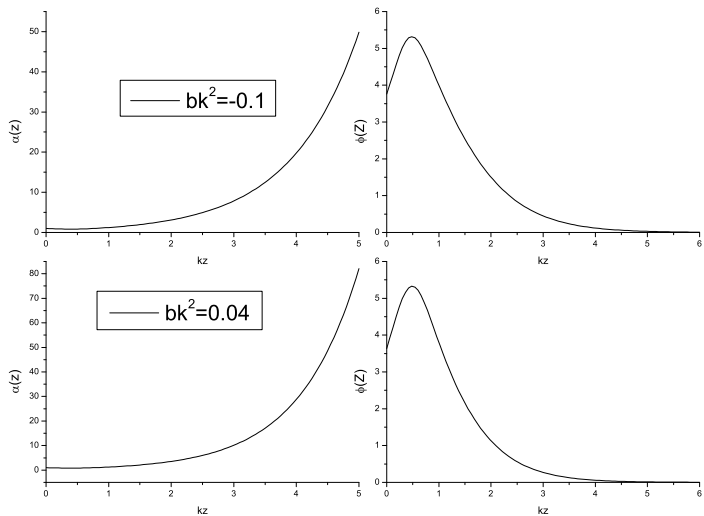}
\end{center}
\caption {The warp factor $a(z)$ and the scalar field $\phi(z)$, as
a function of kz, for $\xi=\xi_c$,  $\hat{\sigma}=5$,
$\hat{\lambda}=0.01$ for two values of $\hat{b}$: $\hat{b}=-0.1$
(upper panel) and $\hat{b}=0.04$ (down panel). The values for
$\sigma$ and $\Lambda$ do not satisfy the fine tuning of the
RS-model.} \label{1}
\end{figure}

In the case of $5D$ conformal coupling $\xi_c=3/16$ and $\hat{b}=0$,
as we have already mentioned in \cite{KF}, the warp factor $a(z)$ is
of the order of unity in a small region near the brane (starts with
$a'(0)$ negative) and increases exponentially ($a(z)\rightarrow e^{k
z}$), as $z\rightarrow +\infty$,  or equivalent the space-time is
$AdS_5$ asymptotically, while the scalar field $\phi(z)$ is nonzero
on the brane (in the region where the warp factor is almost
constant) and tends rapidly to zero for large $z$.

In Fig. \ref{1} we see that the above mentioned behavior for the
warp factor and the scalar field remains, even when the Gauss-Bonnet
term is present. In particular we have plotted the warp factor
$a(z)$ and the scalar field $\phi(z)$, as a function of z, for
$\xi=\xi_c$, for two values of $\hat{b}$: $\hat{b}=-0.1$ and
$\hat{b}=0.04$.

A complete investigation of the spectrum of the solutions of the
model is quite complicated, however we will try to give a
description. We have checked, that for $\hat{b}<0$, the algebraic
system of Eqs. (17) and (18) (junction conditions) has only one real
solution, which corresponds to a warp factor and a scalar field
profile, of the form which is plotted in the upper panel of Fig.
\ref{1}. On the other hand, for $0<\hat{b}\leq \hat{b}_{c3}$
($\hat{b}_{c3}=0.16$), the algebraic system of Eqs. (17) and (18)
has three real solutions which correspond to three static solutions
of the Einstein equations. For $0<\hat{b}<\hat{b}_{c1}$
($\hat{b}_{c1}=0.1227$) the two of them are of the form of the down
panel of Fig. \ref{1} ($Ads_5$ in the bulk), and the third has a
warp factor which vanishes in finite proper distance $z$ and
$\phi(z)$ tends to infinity there (or equivalently it has a naked
singularity in the bulk). In the short interval
$\hat{b}_{c1}<\hat{b}<\hat{b}_{c2}$ ($\hat{b}_{c2}=0.1250$) only one
of the solutions is of the form of the down panel of Fig. \ref{1},
and the other two exhibit naked singularities. For
$\hat{b}_{c2}<\hat{b}\leq0.16$ we have three solutions with naked
singularities. Finally, for $\hat{b}>0.16$ the algebraic system of
Eqs. (17) and (18) has only one real solution, which exhibits a
naked singularity in the bulk.

The above presented analysis is valid for a special choice of the
free parameters of the model ($\xi=\xi_c$, $\hat{\sigma}=5$,
$\hat{\lambda}=0.01$). For a different choice the analysis of the
spectrum of the solutions is quite similar. For example, in the case
of ($\xi=\xi_c$, $\hat{\sigma}=2$, $\hat{\lambda}=0.01$), we obtain
that $\hat{b}_{c1}=0.1247$ and $\hat{b}_{c2}=0.1250$. Note that the
second critical value $\hat{b}_{c2}$ is almost independent from the
value of brane tension $\hat{\sigma}$. For large $\hat{b}$
($\hat{b}>\hat{b}_{c3}$, $\hat{b}_{c3}=1$ )  the model has only one
solution with a naked singularity in the bulk.

In the case of a zero tension brane $\hat{\sigma}=0,
\hat{\lambda}=0.01$ we have also solutions of the form we described
above, with $\phi$ concentrated around zero $(z=0)$ and the
space-time is $AdS_5$ asymptotically. For $0<\hat{b}<0.1250$ we have
three distinct solutions. Two of them are of the form of Fig.
\ref{1} and the third exhibits a naked singularity, with a zero warp
factor in finite distance and infinite scalar field. Note that the
first of the nonsingular solutions is smooth, while the second
exhibits an effectively negative brane tension due to the
discontinuous first derivative of the scalar field for $z=0$.
However, if $\hat{b}$ exceeds the critical value $\hat{b}_c=0.1250$
we obtain three solutions which suffer from naked singularities in
finite proper distance in the bulk with infinite Ricci scalar. Note
that the two critical values $\hat{b}_{c1}$ and $\hat{b}_{c2}$ which
are different in the case of the nonzero tension coincide in the
case of a zero tension brane $\hat{\sigma}=0$, also $\hat{b}_{c3}$
goes to infinity.

\subsection{Class (a) ($\xi<0$)}

\begin{figure}[h]
\begin{center}
\includegraphics[scale=1,angle=0]{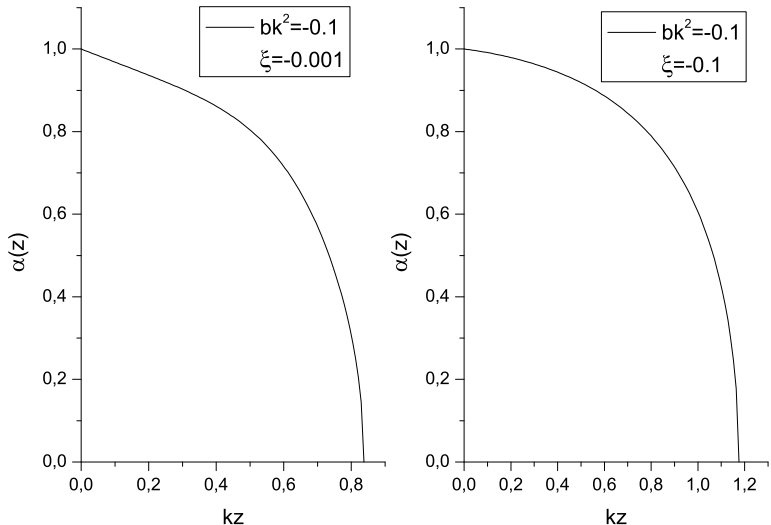}
\end{center}
\caption {The warp factor $a(z)$, as a function of kz, for
$\hat{b}=-0.1$ , $\hat{\sigma}=6$, $\hat{\lambda}=0.01$, for two
values of $\xi$: $\xi=-0.001$ (left-hand panel) and $\xi=-0.1$
(right-hand panel). The values for $\sigma$ and $\Lambda$ satisfy
the fine tuning of the RS-model. } \label{2}
\end{figure}

The solutions for $\xi<0$, as we have seen in our previous paper of
Ref. \cite{KF}, exhibit a naked singularity (in particular the warp
factor becomes zero in finite proper distance in the bulk), while
the scalar field tends to infinity near the singularity.

It is believed that higher order gravity may incorporate quantum
gravity effects which are important when a singularity appears in a
solution of Einstein equations. The question that arises is whether
Gauss-Bonnet gravity can resolve the naked singularities we obtained
in \cite{KF}.

In order to answer the above question, we examined a wide range of
the parameter space of the model, with $\xi<0$, and we found that
Gauss-Bonnet gravity can not overcome the singularity in a
satisfactory way.

A characteristic behavior for the solutions is presented in Fig.
\ref{2}, where we have plotted the warp factor $a(z)$, as a function
of kz, for $\hat{b}=-0.1$ , $\hat{\sigma}=6$, $\hat{\lambda}=0.01$,
and for two values of $\xi$: $\xi=-0.001$ (left-hand panel) and
$\xi=-0.1$ (right-hand panel). We see that naked singularity in the
bulk remains, even when the coupling $\hat{b}$ possesses a nonzero
value. Note that the scalar field tends to infinity near the naked
singularity. The warp factor for $\hat{b}=0$ is presented in Figs.
(1) and (2) in Ref. \cite{KF}. If we compare with the case of
nonzero $\hat{b}$ of Fig. \ref{2} in this work, we see that the
singularity points are only slightly displaced. We have examined
also other values of negative $\hat{b}$, and we observed that for
any negative $\hat{b}$ the solutions are of the form we described
above.

For positive $\hat{b}$ we see that the algebraic system of equations
(13), (17) and (18) has one or more real solutions. Even in this
case we cannot avoid naked singularities in the bulk. However, these
new singularities are of a different nature from the one we
described above. In particular in this case, the second derivative
of the warp factor (or the Ricci scalar) tends to infinity in finite
proper distance in the bulk, while the warp factor, the scalar field
and the first derivative of the warp factor tend to finite values.

For zero tension branes $\hat{\sigma}=0$ we observe the same
behavior with that described above. For $\hat{b}\leq 0$ we have one
singular solution with a warp factor which tends to zero, while for
$\hat{b}>0$ we have only one singular solution with nonzero warp
factor but infinite Ricci scalar.

\subsection{Class (b) ($\xi>\xi_c$)}

\begin{figure}[h]
\begin{center}
\includegraphics[scale=1,angle=0]{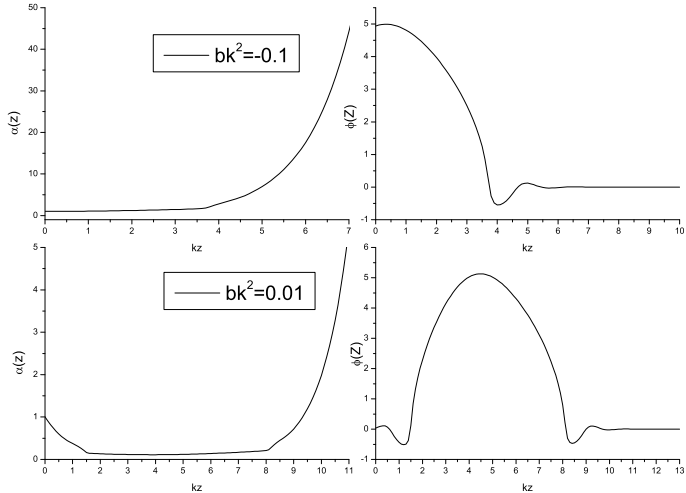}
\end{center}
\caption {The warp factor $a(z)$, as a function of kz, for
$\xi=1+\xi_c$, $\hat{\sigma}=6$, $\hat{\lambda}=0.01$, for two
values of $\hat{b}$: $\hat{b}=-0.1$ (upper panel) and $\hat{b}=0.01$
(down panel). The values for $\sigma$ and $\Lambda$ satisfy the fine
tuning of the RS-model. } \label{3}
\end{figure}

For $\hat{b}=0$ and $\xi>\xi_{c}$, we had obtained a different class
of numerical solutions in Ref. \cite{KF}. In this case the warp
factor $a(z)$ is of the order of unity in a small region near the
brane and increases exponentially ($a(z)\rightarrow e^{k z}$), as
$z\rightarrow +\infty$ (or equivalently the space-time is
asymptotically $AdS_5$), while the scalar field $\phi(z)$ is nonzero
on the brane (in a rather small region near the axes origin where
the warp factor remains constant), and for $kz>>1$ tends rapidly to
zero.

The question that arises is, how these solutions are modified when a
Gauss-Bonnet term is turned on. As we see in the upper panel of Fig.
\ref{3}, for $\hat{b}=-0.1$, the features of the solution remain the
same with those for $\hat{b}=0$. In general, for $\hat{b}<0$, we
have only one static solution of the form we described. We have
checked numerically that for $0<\hat{b}<\hat{b}_{c1}$
($\hat{b}_{c1}=0.0106$) the model possesses two static solutions
which are asymptotically $AdS_5$. The first of them is of the
previously described form, while the second has the same profile
with the first, but it is translated to a finite proper distance in
the bulk, at it is presented in the down panel of Fig. \ref{2}. For
$\hat{b}_{c1}<\hat{b}<\hat{b}_{c2}$ ($\hat{b}_{c2}=0.0192$) we have
also two solutions, but only one is of the form we described above,
while the other exhibits a naked singularity. For
$\hat{b}>\hat{b}_{c2}$ we obtain one or more singular solutions
depending on the value of $\hat{b}$.

\subsection{Class (c) ($0<\xi<\xi_c$)}

\begin{figure}[h]
\begin{center}
\includegraphics[scale=1,angle=0]{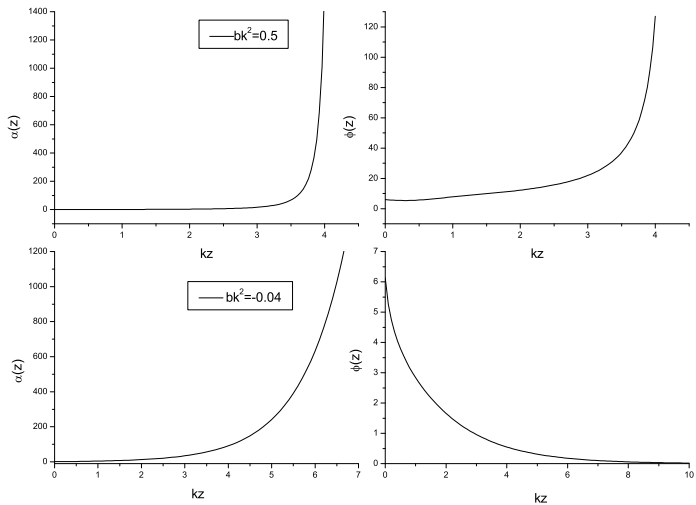}
\end{center}
\caption {The warp factor $a(z)$ and the scalar field $\phi(z)$, as
a function of kz, for $\xi=0.1$, $\hat{\sigma}=6$,
$\hat{\lambda}=0.01$, for two values of $\hat{b}$: $\hat{b}=0.5$
(upper panel) and $\hat{b}=-0.04$ (down panel). In the case of
$\hat{b}=0.5$ we have naked singularity at $z_{s}=0.42$ in the
bulk.} \label{4}
\end{figure}

In the absence of the Gauss-Bonnet term ($\hat{b}=0$) and for
$0<\xi<\xi_{c}$ the warp factor $a(z)$ and the scalar field
$\phi(z)$ tend rapidly to infinity near a singular point in the
bulk, like the upper panel of Fig. \ref{4} (for details see Ref.
\cite{KF}).

We will present an investigation of the spectrum of the static
solutions of the model in the case of a specific choice of the free
parameters of the model $\xi=0.1$, $\hat{\sigma}=6$ and
$\hat{\lambda}=0.01$. For $0<\hat{b}<\hat{b}_{c1}$
($\hat{b}_{c1}=0.0938$) we have two solutions with singular behavior
in the bulk. The first of them exhibits the behavior we described
for $\hat{b}=0$, like the upper panel of Fig. \ref{4}, and the other
has an infinite second derivative of the warp factor, while the
functions $a(z)$, $a'(z)$ and $\phi(z)$ are finite near the singular
point. For $\hat{b}>\hat{b}_{c1}$ we obtain only one solution of the
form the upper panel of Fig. \ref{4}. In the case of negative
$\hat{b}$ and $\hat{b}_{c2}<\hat{b}<0$ ($\hat{b}_{c2}=-0.0485$) we
obtain three distinct solutions: a) Singular solutions where the
warp factor $a(z)$ vanishes in finite proper distance in the bulk,
b) Singular solutions where the second derivative of the warp factor
tends to infinity, while the functions $a(z)$, $a'(z)$ and $\phi(z)$
are finite near the singular point in the bulk and c) Nonsingular
solutions with a warp factor $a(z)$ which increases exponentially
($a(z)\rightarrow e^{k z}$), as $z\rightarrow +\infty$ and a scalar
field $\phi(z)$ which is nonzero on the brane, and for $kz>>1$ tends
rapidly to zero, see the down panel of Fig. \ref{4}. For
$\hat{b}_{c3}<\hat{b}<\hat{b}_{c2}$ ($\hat{b}_{c3}=-0.0536$) we have
three singular solutions, and for $\hat{b}<\hat{b}_{c3}$ we have
only one singular solution.

In general, we observe a similar investigation for other values of
the free parameters of the model. In particular, in the case of
$\hat{\sigma}=0$ we obtain one singular solution of the form of the
upper panel of Fig. \ref{4}, for all the values of the Gauss-Bonnet
coupling $\hat{b}$.

\section{Conclusions and discussion}

We considered brane world models with an additional Gauss-Bonnet
term in the presence of a massless bulk scalar field which interacts
non-minimally with gravity, via a possible interaction term of the
form $-\frac{1}{2} \xi R\phi^2$. We developed a numerical approach
for the solution of the Einstein equations with the scalar field,
and we studied the spectrum of the static solutions, in the case of
a $\lambda \phi^4$ potential, for several values of the free
parameters of the model. We obtained that the existence of the
Gauss-Bonnet term does not change significantly the solutions we
found in our previous work \cite{KF}, in appropriate regions of the
Gauss-Bonnet coupling $\hat{b}$. Outside these regions the solutions
always suffers from naked singularities at finite proper distance in
the bulk. In general the spectrum of the solutions is characterized
from naked singularities except for three cases: i) $\xi>\xi_c$ and
$\hat{b}<\hat{b}_c$ (unstable solutions against scalar field
perturbations with a tachyonic spectrum), ii) $\xi=\xi_c$ and
$\hat{b}<\hat{b}_c$ (stable solutions against scalar field
perturbations with a continuous spectrum of positive energies) and
iii) $0<\xi<\xi_{c}$ and $\hat{b}_{c}<\hat{b}<0$ (stable solutions
against scalar field perturbations with a discrete spectrum of
positive energies). Note, that the third case of stable solutions is
new and does not appear when $\hat{b}$ is equal to zero. Furthermore
these solutions share the same properties: $a(z)$ increases
exponentially with $z$, and tends to an $AdS_5$ space-time
asymptotically, while $\phi(z)$ is nonzero near the brane and
vanishes rapidly when $z$ is increasing. The brane at $z=0$ has a
positive (or zero) brane tension, and the cosmological constant in
the bulk is negative. We note, that in the presence of the
Gauss-Bonnet term we obtain solutions which are $AdS_5$
asymptotically even with positive bulk cosmological constant, but we
have not present our study in this case.

Finally, it would be interesting to discuss if the solutions we
obtained can be used for the construction of a more realistic brane
world model. The main motivation to consider these models was the
observation that the interaction term $-\frac{1}{2}\xi R\phi^2$ can
be interpreted as a mass term, with an effective mass
$m^{2}_{eff}=\xi R$. In the case of the second RS-model we obtain
that $R=16 k \delta(z)-20k^2$, and we see, that for negative $\xi$,
the effective mass term is negative on the brane and positive in the
bulk. This implies an instability of the RS-vacuum near the brane.
As we argued in Ref. \cite{FP}, we expected static stable solutions
with a warp factor similar with that of RS-model, and a scalar field
vacuum with nonzero value on the brane and zero value in the bulk.
These kind of solutions incorporate a layered-phase mechanism for
the localization of Standard-Model particles on the brane. This
mechanism is based on the idea of the construction of a gauge field
model which exhibits a non-confinement phase on the brane and a
confinement phase on the bulk. Thus gauge fields, and more generally
fermions and bosons with gauge charge, can not escape into the bulk
unless we give them energy greater than the mass gap $\Lambda_{G}$,
which emerges from the nonperturbative confining dynamics of the
gauge field model in the bulk. For an extensive discussion see
\cite{FP}.

In \cite{KF} we solved the Einstein equations and we found that
there not solutions with the profile we described above for $\xi<0$.
However we obtained a class of stable solution, for $\xi=\xi_c$
which may be physically interesting. In particular, the warp factor
$a(z)$, for this class of solutions, is of the order of unity near
the brane and increases exponentially ($a(z)\sim e^{k |z|}$ ), as
$z\rightarrow \pm \infty$, while the scalar field $\phi(z)$ is
nonzero on the brane and tends rapidly to zero in the bulk. The
model is completed if we include a second positive tension brane
with $\sigma'/k=6$, in a position $z_c$ where the scalar field
vacuum is practically zero. This two-brane set up is very similar to
the well known RS1-model. In analogy with the RS1-model, we will
assume that the standard model particles live in the effectively
negative tension brane \footnote{The positive tension brane at $z=0$
together with the negative energy density of the scalar field vacuum
$T^{(\phi)}_{00}$, act effectively as a negative tension brane.}, or
visible brane. The advantage of this model is that it incorporates a
layered-phase mechanism for the localization of standard model
particles on the visible brane, and analyzed in detail in Ref.
\cite{KF}.

In this paper we generalize our previous works by considering
brane-models with an additional Gauss-Bonnet term and a bulk scalar
field nonminimally coupled with gravity. As we have already
mentioned the addition of the Gauss-Bonnet term has not a
significant impact to the spectrum of the model, and gives solutions
with the same features as the model without the Gauss-Bonnet term.
However this model exhibits solutions of the interesting form we
described in the previous paragraph, even when $\xi\neq \xi_c$. In
particular these solutions appear when $0<\xi<\xi_{c}$ and
$\hat{b}_{c}<\hat{b}<0$, and can be used for the construction of
realistic brane world models, as we describe above.

We would like to note that in the phenomenological studies in
\cite{Dav} the same model, with an additional Gauss-Bonnet term and
a bulk scalar field nonminimally coupled with gravity, is considered
for $\xi<0$, however the authors of this paper do not take into
account the back reaction of the scalar field to the RS vacuum. We
think that is important for the phenomenological applications to
have a stable background solution, for the complete Einstein
equations and the scalar field.

\section{Acknowledgements} We are grateful to Professors A. Kehagias
and G. Koutsoumbas  for reading and commenting on the manuscript. We
also thank Professors C. Charmousis, I. P. Neupane, E
Papantonopoulos and J. Zanelli for important discussions and
comments. This work is supported by the EPEAEK programme
"Pythagoras" and co-founded by the European-Union
(75$^{\circ}/_{\circ}$) and the Hellenic state
(25$^{\circ}/_{\circ}$).

\end{document}